# THREE-BEAM INSTABILITY IN THE LHC*

A. Burov, FNAL, Batavia, IL 60510, USA

*Abstract*
In the LHC, a transverse instability is regularly observed at 4TeV right after the beta-squeeze, when the beams are separated by about their ten transverse rms sizes [1-3], and only one of the two beams is seen as oscillating. So far only a single hypothesis is consistent with all the observations and basic concepts, one about a third beam - an electron cloud, generated by the two proton beams in the high-beta areas of the interaction regions. The instability results from a combined action of the cloud nonlinear focusing and impedance.

## FACTS AND HYPOTHESES

To prevent transverse instabilities, LHC is normally operated with Landau octupoles and with a damper on [4]. For a single beam in the machine, the octupole instability threshold never exceeded 200A for high chromaticity values, $Q' \geq 10$ and e-fold damping rate 50-200 revolutions [5]. During the recently finished 4TeV proton-proton run, LHC normally worked with maximally available 550A of the octupoles and with full damper gain, but still had regular instabilities at the end of the squeeze [1]. To avoid cancellation of stabilizing beam-beam and octupole anharmonicities [2,6], octupole polarity was switched to positive since summer 2012. As a result, at the end of the squeeze beam-beam nonlinearity effectively provided additional ~200A for the edge ("pacman") bunches and ~400A for regular bunches [6]. At this stage of the process, the edge bunches had 4 times more effective octupole nonlinearity than the single beam threshold, still being unstable. Typically, the instability was observed as intensity loss of the trailing bunches, accompanied with coherent activity at few synchro-betatron lines seen at the BBQ spectrometers.

That high sensitivity of the instability to the beams interaction inclines to suspect coupled-beam oscillations. Indeed, every pacman bunch has 8 long-range beam-beam collisions per interaction region (IR), resulting in $\sim 1.3 \cdot 10^{-3}$ of the incoherent tune shift per every one of the two main interaction regions (IR1 and IR5). This linear tune shift is more than a half of the synchrotron tune, exceeding the rms tune spread on the Landau octupoles at their maximal current of 550A. Although the linear (quadrupolar) parts of incoherent tune shifts at IR1 and IR5 are compensating each other thanks to the crossing horizontal-vertical collision scheme [7], the coherent beam-beam tune shifts are not cancelled, since the two beams have significantly different phase advances between the two interaction points (IP) [8]. Thus, reasons to suspect coupled-beam oscillations as a cause of the end-of-the-squeeze instability seem to be very serious. However, an attentive consideration of these reasons leads to a definite refutation of that suspicion.

First of all, it has to be noted that although the instability is highly sensitive to the presence of both beams in the IR, normally only one of the two beams is seen as unstable (more precisely, only one from the four transverse degrees of freedom is normally seen as unstable). However, this observation does not completely refute the significant coupled-beam contribution to the instability: a role of the apparently stable beam could be hidden by a possible asymmetry of the two-beam oscillations [9,10]. Thus, the apparent stability of one of the beams does not yet refute the coupled-beam hypothesis. This hypothesis is still refuted though, but by another argument, based on the damper consideration.

The LHC transverse damper normally works at rather high gain providing a damping rate of 0.02 inverse revolutions, which is 40% higher than the angular synchrotron frequency $\omega_s$. Originally the damper worked in a narrow-band regime with FWHM of its time-domain response ~140ns, so high frequency coupled-bunch modes of 50ns beams were not effectively damped. Last several months of the Run I the damper worked in a broadband, really a bunch-by-bunch regime [11], but that did not show any improvement for the instability. That new bunch-by-bunch damper is broadband enough to resolve coherent motion of every bunch, but it cannot resolve intra-bunch motion; it sees only a centroid of every individual bunch, thus reacting to every head-tail mode proportionally to a weight of the centroid in its oscillations. At a sufficiently high damper gain, this means that only those modes are unstable which have practically zero centre of mass amplitude. These modes are invisible for the damper and thus can be unstable due to the machine impedance. It is important that beam-beam coupling for that sort of potentially unstable modes is suppressed by the same reason as their visibility for the damper. Indeed, for the long-range collisions, the bunch length is much smaller than the beta-functions, so kicks of the oncoming bunches are equivalent to kicks of their centres. Since the bunch centres are blocked by the damper, the beam-beam coupling is strongly suppressed, so beam-coupling cannot play a significant role. This qualitative refutation of the coupled-beam contribution in case of a strong damper can be expressed by means of a simple model treating coupling of two head-tail modes of the two beams.

Let $A_{1,2}$ be amplitudes of the eigenmodes in beam 1 and beam 2. Due to the beam-beam interaction, they become coupled. Assuming for simplicity a single IP, the mode dynamic equations follow:

$$\dot{A}_1 = -i\omega_c \dot{A}_1 - d\alpha A_1 - iq\alpha A_2 ;$$
$$\dot{A}_2 = -i\omega_c \dot{A}_2 - d\alpha A_2 - iq\alpha A_1 .$$

Here $\omega_c$ is the impedance-related coherent tune shift of the separated beams; the parameter $\alpha$ reflects a weight of the centre of mass in the amplitudes $A$ so that at zero chromaticity $\alpha = 1$ for the $0^{th}$ head-tail mode; $d$ and $q$ are the damping rate and beam-beam tune shift. A straightforward solution shows that this system has two coupled modes (so called $\pi$ and $\Sigma$ modes) with frequencies

$$\Omega_\pm = \omega_c - id\alpha \pm q\alpha .$$

To be unstable and thus require some Landau damping to stabilize it, the mode centre of mass parameter has to be small enough: $\alpha < \text{Im}\,\omega_c / d$. From here, the coupled-beam tune shift is limited as

$$q\alpha < \text{Im}(\omega_c)q/d.$$

When the gain $d$ is high enough, the beam-coupling correction just slightly shifts the coherent tunes, so that their position in the stability diagram remains almost the same. In case when the beam-beam octupolar term adds up to the Landau octupoles, the stability diagram increases, so that the two beams are more stable than one. For LHC at the end of the beta-squeeze, the beam-beam tune shift per IR and the damping rate are close to each other, $q/d\sim 1$, so the coupled-beam tune shift is limited as $q\alpha < \text{Im}(\omega_c)$. Thus, in this case, the beam-beam coupling moves coherent tune shifts along their real axis by a value not exceeding their imaginary part. However, the stability diagram width (say, FWHM) is 3-10 times higher than its height; moreover, with the damper, imaginary parts of the coherent tunes are much smaller than their real parts [12], so a shift of the real parts of the coherent tunes at the value limited by its imaginary part results only in a small increase of the required octupolar current, in any case smaller than ~30%, and much smaller than that for the LHC impedance model. Taking into account that beam-beam octupolar term increases the stability diagram at least by 40%, it can be concluded that the two beams have to be more stable than one – in a contradiction to the observations. Thus, the effect of coupling oscillations of the two beams cannot explain the observed instability at the end of the squeeze.

For those who may be not quite convinced by the qualitative explanation and the model above, suspecting them to be over-simplified, the author provided a detailed solution of Vlasov equation, where the azimuthal, radial, coupled-bunch, and coupled-beam mode dimensions were taken into account in a framework of the Nested Head-Tail (NHT) Vlasov solver [12]. The result of that detailed computation confirmed the conclusions above: two-beam stability requires almost the same stabilizing octupolarity as a single beam does; with the beam-beam octupolar term taken into account it means the two beams have to be stable at less than 100A of the Landau octupoles, while in reality they are not stable even at the maximally available 550A. Almost at the same time similar result was obtained by S. White for single-bunch beam-beam tracking simulation with Beam-Beam3D program [13]. According to his results, stability conditions for weak-strong and strong-strong collisions are almost the same when the damper is fully on.

To verify these considerations, a special LHC beam experiment was run, where two beams with 78 bunches each were able to see or not see each other in the interaction regions by means of RF cogging ("cogging MD"). On top of that, tune separations of the two beams were varied up to several times of the beam-beam tune shift per IR [14]. Despite a relatively small number of bunches (78⊗78), the end-of-squeeze instability was still observed. It was seen that the instability is not sensitive to that large tune separation, while it is sensitive to simultaneous presence of the two beams in the IR1 and IR5 [15]. Thus, the three-level theoretical refutation of the coupled-beam oscillations as a cause of the instability was supported by its experimental refutation. Then, what is the cause of the instability?

Well, the fact is that when a reference beam sees another beam in the IR, it is much more unstable. The other beam, being rock-stable, dramatically changes life conditions of the reference beam. The Coulomb field of the other beam makes the reference beam even more stable than it would be alone. Hence, the other beam brings with itself something else, a third element, which interacting with the reference beam makes the beam much more unstable. What can that third element, created by the two beams in the IR, be?

This third element cannot be a high order mode (HOM) electromagnetic field excited by joint efforts of the two beams inside a parasitic cavity located somewhere in IR. Indeed, that sort of coherent tune shift for two beams cannot be higher than a doubled tune shift of a single beam. Moreover, the two-beam HOM-driven tune shift is coming closer to the doubled single beam tune shift only if the dominant part of the entire single beam tune shift is driven by that HOM, which cannot be the case since the observed instability for 78⊗78 bunches does not show any difference from 1378⊗1378 bunches. At the same time, while the single beam is stabilized by 200A, the two beams are unstable with 550+200=750A of the effective octupole current. That is why the sought-for third element cannot be a HOM of one or another parasitic cavity in the IR, it cannot be a free EM field. If this third element is not an EM field, it can be only matter, attracted by the two beams in the IR and disappearing when one of the beams is not there. It appears to be very clear that this matter can be nothing but an electron cloud in the IR.

## E-CLOUD AS NONLINEAR LENSE

Electron cloud influences proton oscillations in two aspects.

First, it works as a static lens, shifting up all coherent and incoherent tunes. This lens is nonlinear; the tune shifts of the transverse tails should be smaller than those of the core. Nonlinearity of this lens changes the proton stability diagram. The second aspect is that e-cloud is a reactive medium, whose response to proton perturbations is similar to a low-Q impedance [16-18]. Impedance of the electron cloud moves coherent tune shifts of the proton beam.

Electron cloud is not homogeneous along the bunch length; its line density changes and it may have multiple transverse pinches, so accurate computation of its effect on the proton coherent motion is very complicated. So far approaches in this direction are based either on simplified analytical models [16-18] or heavy multi-particle tracking [19,20]. Below, both focusing (static) and reactive (dynamic) aspects of the electron cloud are taken into account within a framework of a simplified model, where the cloud is represented as a longitudinally homogeneous electron density distribution, or a beam with zero longitudinal velocity, whose transverse profile is identical to one of the Gaussian proton beam. It can be rephrased that only electrons within the transverse radius of the proton beam are taken into account, while all the outside parts of the cloud are neglected both for the focusing and impedance aspects.

With $N_e$ electrons along the entire LHC circumference seen by the proton beam of the normalized rms emittance $\varepsilon_n = \gamma \varepsilon$, the incoherent proton tune shift on the electrons $\Delta_e Q_x$ can be expanded over the proton actions $J_x, J_y$ [21]:

$$\Delta_e Q_x = \Delta_e Q_x^{(0)} + \Delta_e Q_x^{(1)} + ...$$

$$\Delta_e Q_x^{(0)} = \frac{N_e r_p}{4\pi \varepsilon_n};$$

$$\Delta_e Q_x^{(1)} = -\frac{3 \Delta_e Q_x^{(0)}}{8} \frac{J_x + 2J_y/3}{\varepsilon}; \quad \langle J_{x,y} \rangle = \varepsilon.$$

In the weak head-tail approximation, the eigenvalues $Q$ are to be found as solutions of the dispersion equation [22]

$$1 = -Q_c \int \frac{J_x \partial F / \partial J_x}{Q - lQ_s - \Delta Q_x + io} d\Gamma,$$

where $F$ is a normalized phase space density defined on the phase space $\Gamma$, $Q_c$ is the coherent tune shift, which gives the mode tune in case of no tune spread $\Delta Q_x$, $Q_s$ is the synchrotron tune, $l$ – azimuthal mode number and $o$ - infinitesimally small positive value. The stability diagram is a map of the real axis in a complex plane Q onto a complex plane

$$D = \left( -\int \frac{J_x \partial F / \partial J_x}{Q - lQ_s - \Delta Q_x + io} d\Gamma \right)^{-1},$$

so the mode is stable if and only if its tune shift $Q_c$ is located inside the stability diagram. For Gaussian transverse distribution, and with negligible spread of the synchrotron frequencies, the 2D dispersion integral was found by R. Gluckstern [22]:

$$\int_0^\infty \int_0^\infty \frac{x \exp(-x-y) dx dy}{q - ax - by + io} =$$
$$-\frac{a - b + (b - q(1 - b/a)) \exp(-q/a) \mathrm{Ei}(q/a)}{(a-b)^2} +$$
$$\frac{b \exp(-q/b) \mathrm{Ei}(q/b)}{(a-b)^2} -$$
$$\pi i \frac{|-(b - q(1 - b/a)) \exp(-q/a) \theta(q/a) + b \exp(-q/b) \theta(q/b)|}{(a-b)^2};$$

$$\mathrm{Ei}(z) \equiv -P.V. \int_{-z}^\infty (e^{-t}/t) dt.$$

Here *P.V.* stays for the principle value and $\theta(z)$ is the Heaviside theta-function. Stability diagrams for distribution functions $F(J_x, J_y) \propto (1 - (J_x + J_y)/a)^n$ are discussed in Ref. [23].

The incoherent tune shift $\Delta Q_x$ in the denominator of the dispersion integral takes into account all the nonlinearities: Landau octupoles, beam-beam, e-cloud, and the remaining machine nonlinearities if they cannot be neglected: $\Delta Q_x = \Delta_o Q_x + \Delta_{bb} Q_x + \Delta_e Q_x + .....$ The octupoles incoherent tune shift contribution is described by a symmetric matrix [24]:

$$\begin{pmatrix} \Delta_o Q_x \\ \Delta_o Q_y \end{pmatrix} = \begin{pmatrix} a_o & b_o \\ b_o & a_o \end{pmatrix} \begin{pmatrix} J_x / \varepsilon \\ J_y / \varepsilon \end{pmatrix};$$

for the normalized rms emittance $\varepsilon_n = 2\mu m$ and octupole current $I_o = +100 A$, the LHC octupole matrix elements were computed as [24]

$$a_o = 4.2 \cdot 10^{-5}; \quad b_o = -2.9 \cdot 10^{-5}$$

at 4TeV.

Approximating the interaction region as a drift space, the long-range beam-beam octupole contribution per IR is computed as

$$\Delta_{bb} Q_x^{(1)} = \frac{3 |\Delta_{bb} Q_x^{(0)}|}{2r^2} \frac{J_x - 2J_y}{\varepsilon}$$

with $\Delta_{bb}Q_x^{(0)}$ as the quadrupole beam-beam tune shift per IR, and $r$ as the normalized beam-beam separation, or the separation in the units of rms beam sizes, which is almost the same for all the long-range collisions. At the end of the squeeze, $|\Delta_{bb}Q_x^{(0)}| = 2.5 \cdot 10^{-3}$, $r = 9.5$.

One of the main issues associated with multiple contributions to the incoherent tune shift $\Delta Q_x$ is a possibility of significant reduction of the stability diagram. When it was realized that the Landau octupoles and beam-beam contributions may almost cancel each other for negative octupole polarity [Stephane, Xavier], their polarity was inverted. For positive octupole polarity, these two contributions add together. According to the LHC impedance model [Nicolas], the coherent tune shifts of unstable modes are all negative [Burov]. At the left (defocusing) side of the stability diagram, the beam-beam contribution at the end of the squeeze is approximately equivalent to 200A for pacman bunches.

Electron cloud may significantly change the stability diagram: defocusing anharmonicity of the cloud may almost cancel common focusing anharmonicity of the octupoles and beam-beam, resulting in a collapse of the focusing side of the diagram. The tune shifts formulas above show that at the end of the squeeze with 500A of the Landau octupoles this requires $N_e \simeq 1 \cdot 10^{10}$ seen by the proton beam within its size along the entire orbit. This collapse of the focusing part of the stability diagram would not yet lead to instability, were the coherent tune shifts of unstable modes all negative, as they are computed [12] for the LHC impedance model [25]. However, the electron cloud not only changes the stability diagram, it also introduces its own impedance. Tune shifts of unstable modes driven by this impedance are mostly positive.

## IMPEDANCE OF E-CLOUD

Electron cloud is a dynamic object: it responds to collective perturbations of the proton bunches. Being excited by these perturbations, a dipole moment of the cloud oscillates, then, in the proton Coulomb field. Due to significant nonlinearity of this field, the excited electron perturbation has a high frequency spread and decoheres rather fast. This consideration leads to an idea to represent the cloud coherent response by means of a resonator wake function with rather small $Q$-factor, $Q \sim 2\text{-}5$. [16-18]. To estimate this wake function, the proton bunch can be substituted by a piece of a coasting beam with constant 3D density, equal to an average density of a Gaussian bunch

$$\bar{n}_p = N_b^{-1} \int n_p^2(\mathbf{r}) d\mathbf{r} = \frac{N_b}{8\pi^{3/2} \sigma_x^2 \sigma_z}.$$

In the Coulomb field of this homogeneous bunch, electrons oscillate with an angular frequency

$$\bar{\omega}_e = c\sqrt{2\pi \bar{n}_p r_e} = c\sqrt{\frac{N_b r_e}{4\sqrt{\pi} \sigma_x^2 \sigma_z}}.$$

Let a small longitudinal sample of this bunch have a charge $q$ and a rigid offset $x_p$. Due to its dipole moment $qx_p$, this proton sample excites an electron velocity

$$v_e = \frac{qex_p}{2\sigma_x^2 mc},$$

leading to an amplitude of the electron offset $x_e = v_e / \bar{\omega}_e$. Modelling the electron beam by the transversely homogeneous one, same as the proton one, the kick to the protons is calculated. This kick can be expressed in terms of the cloud wake function; using the same convention as in Ref.[Chao], this yields ($\tau \leq 0$):

$$W_e(\tau) = c\frac{R_S}{Q} = \frac{N_e r_e}{4\sigma_x^4} \frac{c}{\bar{\omega}_e} \sin(\bar{\omega}_e \tau) \exp(\bar{\omega}_e \tau / 2Q),$$

where $N_e$ is the total number of electrons seen by the proton bunch at the given part of the orbit. Note that sign of this wake is the same as for the conventional cavity modes: its derivative is positive at $\tau = -0$. This wake differs only by a factor of $\pi^{1/4} \approx 1.3$ from one suggested in Ref. [17] what appears to be well within error bars of both derivations.

Coherent tune shifts caused by the electron cloud wake field can be estimated within the air-bag approximation. Neglecting bunch coupling and assuming the weak head-tail approximation, the coherent tune shift can be presented as in Eq. (6.188) of Ref. [26]:

$$Q_c = -i\frac{N_b r_p \beta_x}{8\pi^2} \int_{-\infty}^{\infty} Z_x(\omega) J_l^2(\omega \tau_b - \chi) d\omega;$$

$$Z_x(\omega) = \frac{c}{\omega} \frac{R_S}{1 - iQ\left(\frac{\omega}{\bar{\omega}_e} - \frac{\bar{\omega}_e}{\omega}\right)}.$$

Here $\beta_x$ is the beta-function assumed to be weighted with the impedance $Z_x$ along the orbit, $\tau_b \simeq \sqrt{2}\sigma_z / c$ stands for the air-bag equivalent bunch length, and $\chi = Q'_x \omega_0 \tau_b / \eta$ is the conventional head-tail phase with $Q'_x \equiv p dQ_x / dp$ as a chromaticity, $\omega_0$ as the angular revolution frequency and $\eta$ as the slippage factor. Substitution of the cloud impedance into the air-bag formulas for $\chi \lesssim 1$ yields:

$$\operatorname{Im} Q_c = \chi \Delta_e Q_x^{(0)} F_R ;$$
$$\operatorname{Re} Q_c = \Delta_e Q_x^{(0)} F_I ;$$
$$F_R = \sqrt{\frac{2}{\pi}} Q \phi_e \int_0^\infty \frac{J_l(x\phi_e) J_l'(x\phi_e)}{1 + Q^2 (x - 1/x)^2} \frac{dx}{x} ;$$
$$F_I = \frac{Q^2 \phi_e}{\sqrt{2\pi}} \int_0^\infty \frac{J_l^2(x\phi_e)(x - 1/x)}{1 + Q^2 (x - 1/x)^2} \frac{dx}{x} .$$

Here $\phi_e = \bar{\omega}_e \tau_b = \sqrt{2} \bar{\omega}_e \sigma_z / c$ is a phase advance of the electron oscillations on the air-bag bunch length $\sqrt{2}\sigma_z$. According to Ref. [17], for round beams $Q \simeq 5$. For this Q-factor, the resonator impedance form-factors $F_R, F_I$ as functions of the phase advance $\phi_e$ are presented in Fig. 1,2

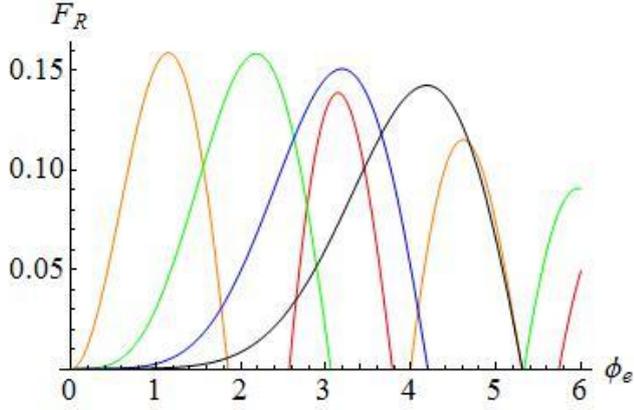

Fig. 1: Growth rate formfactor $F_R$ for head-tail modes 0-4 (consequently red, orange, green, blue and black curves).

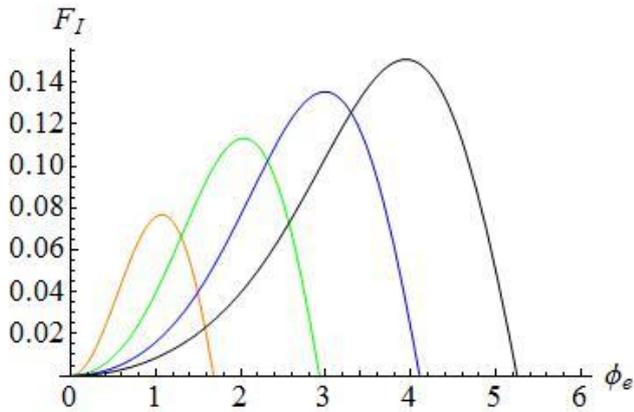

Fig. 2: The same for the tune shift formfactor $F_I$.

As it is seen from the results above, the growth rate of the most unstable head-tail mode $\max_l(\operatorname{Im} Q_c)$ is almost independent of the beta-function, at least directly, since the incoherent tune shift $\Delta_e Q_x^{(0)}$ does not contain any explicit dependence on that, and the formfactor $F_R$ of the most unstable mode is almost constant. Certain dependence on the beta-function is implicitly contained in the tune shift $\Delta_e Q_x^{(0)}$ due to some sensitivity of the e-cloud build-up to the beam size, but this issue is beyond a scope of this paper. It is already clearly seen that the head-tail number of the most unstable mode $l_*$ is about equal to the integer part of the phase $l_* \simeq \phi_e \propto 1/\sqrt{\beta_x}$. For the LHC, the orbit-average $\beta_x = R_0/Q_x \approx 70$m yields the phase advance $\phi_e = 20$rad and thus the same number of the most unstable mode, $l_* \simeq 20$. In the reality those high-order head-tail modes should be suppressed by a spread of the synchrotron tunes. That is why a possible e-cloud accumulation inside the regular part of the machine contributes to the Landau damping, while its contribution to the effective impedance can be neglected. The situation dramatically changes at the end of the squeeze, when beta-functions reach a level of few km for significant part of the interaction regions. For instance, at $\beta_x = 4$km, the phase advance $\phi_e \simeq 2$rad, and so the head-tail number is not that high: $l_* = 2$.

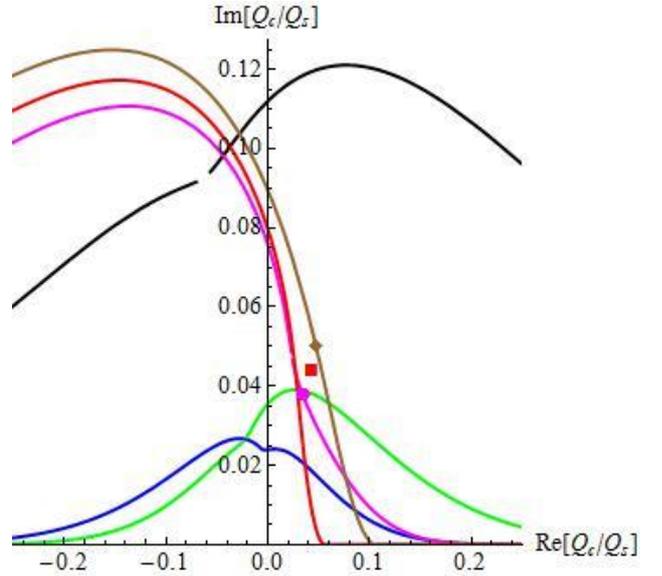

Fig. 3: LHC stability diagrams: a separated stable beam with +200A of the Landau octupoles (green); pacman beam-beam only (no octupoles) at the end of the squeeze (blue); this pacman beam-beam and +500A of the octupoles in addition (black); same as the black line plus e-cloud with total $N_e = (1.3, 1.5, 1.7) \cdot 10^{10}$ (magenta, red, brown). Markers of the corresponding colour show the most unstable modes.

In the Fig. 3, several LHC stability diagrams are shown together with the coherent tune shifts of the most unstable modes. Several important aspects of this figure deserve to be discussed.

1. According to Fig. 3, the instability happens if and only if the total number of electrons belongs to a certain interval: $1.3 \cdot 10^{10} \leq N_e \leq 1.7 \cdot 10^{10}$. This may raise a suspicion that this instability can hardly happen since it requires a rather narrow interval of the cloud intensities. However, this suspicion can be counter-argued that the upper limit of the instability may not be so important. Indeed, as soon as the electron population reaches the lower instability threshold, the instability itself may prevent further accumulation of the electrons, and thus the cloud intensity will never reach the upper instability threshold. Still, the instability may stop due to emittance growth and intensity loss of the proton beam, caused by the instability itself. That sort of scenario appears to be consistent with observations.
2. While the collapse of the right (focusing) side of the stability diagram is driven by the total number of electrons seen by the beam along the orbit, the coherent tune shifts of the unstable modes are driven to the right by the electrons seen at high-beta (~ km-range) areas only. Fig. 3 does not make any difference between these two groups of electrons; in other words, it assumes that all the electrons are mainly accumulated in the high-beta areas. If the opposite is true, the right-side collapse of the stability diagram would not lead to the instability: the electron impedance does not play a role in that case, while all the coherent tune shifts of the unstable modes are negative [12] according to the currently accepted impedance model [25].
3. However, the LHC impedance model is not so certain. Measured single-beam thresholds and single-bunch tune shifts are consistent with 2-3 times higher impedance at the single-bunch (~GHz) frequency range than it is calculated in Ref. [7, 25]. An origin of this discrepancy is so far unknown. In case this lost impedance is mostly associated with a broadband resonator, underestimated in the computations, the impedance-related unstable coherent tune shifts will appear at the focusing part of the stability diagram, and a smaller value of the e-cloud impedance would be sufficient to explain the instability. In that case the fraction of the e-cloud in the interaction region may be smaller or even much smaller than the contribution of the regular part of the orbit. One more reason for reduction of the threshold electron population in the high-beta parts of the IRs can be found in Ref. [27,28] suggesting significant enhancement factor for the cloud wake function.
4. It has been mentioned above that the head-tail number of the most unstable mode depends on the beta-function of the cloud localization. For the average beta-function in the LHC, about 70m, this number is very high, $l_* \simeq 20$, so these modes should be stabilized by the spread of the synchrotron tunes, entering as $l_* \Delta Q_s$. However, during the ramp and then at the flat top the bunch length is reduced, and so is the synchrotron tune spread. On top of that, some e-clouds could be accumulated at the areas of maximal beta-functions of the regular cells, where $\beta_{x,y} \simeq 200\text{m}$, and thus $l_* \simeq 10$. Maybe, due to the ramp these modes are not suppressed any more by the longitudinal Landau damping, and thus become unstable. Their instability cannot be seen by BBQ spectrometers since the bunch oscillations are too microwaving, at the ~10GHz frequency range. Instability of these microwave modes could be an explanation for the emittance growth at the LHC ramp and losses during and after that [29, 30].
5. Computations of this paper neglect the damper. Excitation of the microwave modes $l_* > 2$ should not be sensitive to the damper seeing the bunch centre only.

## SUMMARY

Accumulation of an electron cloud in the high-beta areas of the ATLAS and CMS interaction regions so far appears to be the only acceptable hypothesis explaining the transverse instability at the end of the beta-squeeze in the LHC. According to that hypothesis the instability develops due to two different effects of the e-cloud: collapse of the focusing side of the stability diagram and introduction of the broadband impedance at GHz frequency range at the end of the squeeze. The purpose of this paper was to show that this hypothesis is compatible with all known observations and main conventional ideas.

Finally, I would like to stress that all computations of this paper are extremely approximate, with unknown error bars. An electron cloud model applied above is very simplified; many other factors are absolutely neglected - the bunch-by-bunch damper, radial head-tail modes, couple-bunch interaction. Certainly all these factors require more detailed and thorough future analysis.

## ACKNOWLEDGMENT

This paper would not be possible without numerous discussions with many of my CERN colleagues that happened during my visit to CERN as a FNAL-LARP long-term visitor. I am especially thankful to S. Fartoukh, E. Metral, N. Mounet, E. Shaposhnikova and S. White for sharing of observations, ideas and results about instabilities in the LHC.

# REFERENCES

[1] E. Metral, "Review of the instabilities", talk at the LHC Beam Operation Workshop, Evian, 2012.

[2]. X. Buffat, "Instability observations", talk at CERN LBOC meeting, Aug. 14, 2012.

[3] G. Arduini, "Summary on observations on the instabilities", talk at CERN LMC meeting Aug 15, 2012.

[4] W. Hofle, "Transverse damper", Proc. LHC Performance Workshop, Chamonix, 2012.

[5] N. Mounet, "Beam stability with separated beams", talk at the LHC Beam Operation Workshop, Evian, 2012.

[6] S. Fartoukh, "On the sign of the LHC octupoles", CERN LMC talk, July 11, 2012.

[7] LHC Design Report, Volume 1, CERN-2004-003-V-1, 2004.

[8] S. Fartoukh, private communication, 2012.

[9] V. Lebedev, private communication, 2012.

[10] A. Burov, "Two-beam instability in electron cooling", Phys. Rev. ST Accel. Beams **9**, 120101.

[11] D. Valuch, "Transverse damper system", talk at the LHC Beam Operation Workshop, Evian, 2012.

[12] A. Burov, "ADT suppression of coherent beam-beam", ICE meeting, CERN, Oct. 30, 2012; "Nested Head-Tail Vlasov Solver", CERN AP Forum talk, Dec 4, 2012.

[13] S. White, "Update on long-range instabilities", ICE section meeting, CERN, Nov. 7, 2012.

[14] S. Fartoukh, "Two-beam impedance MD request", CERN LSWG meeting, Sep. 24, 2012.

[15] S. Fartoukh, E. Metral, A. Burov, conclusions of the cogging MD, Dec. 2012.

[16] A. Burov, N. Dikansky, "Electron Cloud Instabilities", Proc. of Int. Workshop on Multibunch Instabilities in Future Electron and Positron Accelerators, Tsukuba, KEK, 1997.

[17] K. Ohmi and F. Zimmermann, "Head-Tail Instability Caused by Electron Clouds in Positron Storage Rings", Phys. Rev. Lett. 85, 3821–3824, 2000.

[18] K. Ohmi, F. Zimmermann, E. Perevedentsev, "Wake-field and fast head-tail instability caused by an electron cloud", Phys. Rev. E 65, 016502, 2001.

[19] K. Li, "Instabilities Simulations with Wideband Feedback Systems", Proc. ECLOUD'12 workshop, La Biodola, Italy, 2012.

[20] H. Bartosik, "Benchmarking of Instability Simulations at LHC", ibid.

[21] A. Burov, V. Lebedev, "Transverse instabilities of coasting beams with space charge", Phys. Rev. ST Accel. Beams 12, 034201 (2009).

[22] J. S. Berg, F. Ruggiero, "Landau damping with two-dimensional tune spread", CERN-SL-AP-96-071-AP, 1996.

[23] E. Metral, A. Verdier, "Stability diagram for Landau damping with a beam collimated at an arbitrary number of sigmas", CERN-AB-2004-019-ABP, 2004.

[24] E. Metral, N. Mounet, private communications, 2012.

[25] N. Mounet, "The LHC Transverse Coupled-Bunch Instability" PhD Thesis, EPFL, Lausanne, 2012.

[26] A. Chao, "Physics of Collective Beam Instabilities in High Energy Accelerators", 1993.

[27] F. Zimmermann, "Review of single bunch instabilities driven by an electron cloud", Phys. Rev. ST Accel. Beams 7, 124801, 2004.

[28] E. Benedetto, D. Schulte, F. Zimmermann, and G. Rumolo, "Simulation study of electron cloud induced instabilities and emittance growth for the CERN Large Hadron Collider proton beam", Phys. Rev. ST Accel. Beams 8, 124402, 2005.

[29] F. Roncarolo, "Beam size measurements", LHC Beam Operation Workshop, Evian, 2012.

[30] G. Papotti, "Beam losses through the cycle", ibid.